# A Data-Driven Framework for Designing Microstructure of Multifunctional Composites with Deep-Learned Diffusion-Based Generative Models

Kang-Hyun Lee[1,1)], Hyoung Jun Lim[1,1)] and Gun Jin Yun[2,1),2)]

[1)] *Department of Aerospace Engineering, Seoul National University, Seoul, South Korea*
[2)] *Institute of Advanced Aerospace Technology, Seoul National University, Gwanak-gu Gwanak-ro 1, Seoul 08826, South Korea*

## ABSTRACT

This paper puts forward an integrated microstructure design methodology that replaces the common existing design approaches: 1) reconstruction of microstructures, 2) analyzing and quantifying material properties, and 3) inverse design of materials using deep-learned generative and surrogate models. The long-standing issue of microstructure reconstruction is well addressed in this study using a new class of state-of-the-art generative model, the diffusion-based generative model (DGM). Moreover, the conditional formulation of DGM for guidance to the embedded desired material properties with a transformer-based attention mechanism enables the inverse design of multifunctional composites. A convolutional neural network (CNN)-based surrogate model is utilized to analyze the nonlinear material behavior to facilitate the prediction of material properties for building microstructure-property linkages. Combined, these generative and surrogate models enable large data processing and database construction that is often not affordable with resource-intensive finite element method (FEM)-based direct numerical simulation (DNS) and iterative reconstruction methods. An example case is presented to demonstrate the effectiveness of the proposed approach, which is designing mechanoluminescence (ML) particulate composites made of europium and dysprosium ions. The results show that the inversely-designed multiple ML microstructure candidates with the proposed generative and surrogate models meet the multiple design requirements (e.g., volume fraction, elastic constant, and light sensitivity). The evaluation of the generated samples' quality and the surrogate models' performance using appropriate metrics are also included. This assessment demonstrates that the proposed integrated methodology offers an end-to-end solution for practical material design applications.

**Keywords**: Deep learning; Diffusion-based generative models; Particulated composite; Microstructure; Mechanoluminescence; Data-driven inverse design

---

[1] These authors have constributed equally to this work
[2] Corresponding author, Professor, Department of Aerospace Engineering, Seoul National University, Gwanak-gu Gwanak-ro 1, Building 301, Room 1308, Seoul, 08826, South Korea, Tel) +82-2-880-8302, Email) gunjin.yun@snu.ac.kr

# 1. Introduction

Hybrid multifunctional composite materials represent a class of advanced materials with multiple functionalities such as structural strength, thermal conductivity, electromagnetic shielding, and sensing capabilities [1-3]. These materials typically consisted of various components, including fibers, resins, and coatings, engineered to achieve the desired properties synergistically. As establishing the process-structure-property (PSP) for multifunctional composite materials has been a prominent research area, it is imperative to develop methods for analyzing the interrelationships between properties and microstructures on a multiscale basis. This approach moves beyond traditional work that relies on empirical evaluation and trial and error [4-6]. In this context, the integrated computational materials engineering (ICME) strategy proposed by the National Materials Advisory Board (NMBD) is being applied as a core technology for the development of advanced multifunctional materials [7-9].

In particular, computational materials engineering using numerical techniques such as the finite element method (FEM) along with microstructure characterization and reconstruction (MCR) algorithms [10-15] is becoming a crucial tool for understanding material behavior and performance in various applications [16, 17]. For instance, determining the representative volume element (RVE) is often required for understanding the material's microstructure computational analysis of composite material behavior [18-22]. If the RVE is defined, we can predict how the material will behave under different loads and environmental conditions with multiscale analysis. Among various experimental inspection technologies for microstructure characterization for RVE definition (such as optical microscopy, scanning electron microscopy (SEM) [23], transmission electron microscopy (TEM), and X-ray diffraction (XRD) analysis [24]), micro-computed tomography (micro-CT) is the most powerful tool that allows for visualizing the internal structure of a material in a 3D domain. It has been used to identify the constituents in the microstructure, including the shape, size, and distribution of the matrix and reinforcement and any defects or discontinuities present in the material [25]. In addition, it is beneficial for analyzing the behavior of composite structures with 3D FEM. For example, Kim and Yun [26] utilized X-ray CT images of particulate composites for constructing high-fidelity FEM models. They presented the correlation of morphological descriptors of microstructure and stress using FEM-based direct numerical simulation (DNS). However, FEM-based composite modeling and analysis consume significant computational resources. Moreover, due to the heterogeneous nature of the fillers, data size, and challenging image processing, it is formidable to sample subscale structure over a large area, which leads to micro-CT being time-consuming, costly, and proficiency-dependent. Therefore, many works focus on the implementation of a reconstruction algorithm for microstructure generation following the physical features of materials. [10-14, 27]. Moreover, the FEM analysis is unsuitable for design and optimization problems as computationally intensive analyses are conducted iteratively. As a result, reduced order models (ROMs) are introduced to decrease computational costs while maintaining high accuracy and versatility [28]. Some notable ROMs are based on the principle component analysis (PCA) [29], proper generalized decomposition (PGD) [30], fast Fourier transformation (FFT) [31], transformation filed analysis (TFA) [32], nonuniform transformation field analysis (NTFA) [33], proper orthogonal decomposition (POD) [34], and self-consistent clustering analysis (SCA) [35].

In addition to this subject, there has been a growing trend to use deep learning (DL) models for material design in recent years [36-38]. The primary application of deep learning (DL) is to unveil the intricate relationships between high-dimensional data (e.g., such as microstructure geometry) and the specific variables of interest (e.g., microstructural descriptors or material response). It is achieved by leveraging established experimental or simulated databases. For example, Yang et al. [39] investigated the stress-strain prediction depending on the binary microstructure image using DL models to predict the composite failures beyond the elastic limit. Stress-strain curves are chosen as the model's result because they are difficult to predict, given their high dimensionality. This approach is conducted in many fields to reduce the computational resources for composite microstructures and lattice structures, protein structures, and origami/kirigami structures [40-42]. Beyond the prediction of stress-strain response, there have been numerous efforts to investigate the use of DL models for generating microstructures that have specific desired morphology [43-51], as well as models that can predict mechanical and thermal behavior based on the microstructure geometry [46, 52].

In particular, generative models such as variational autoencoders (VAEs) and generative adversarial networks (GANs) have been extensively studied to explore design space for various microstructures. A notable characteristic of these generative models is that they learn a compressed (i.e., low-dimensional) representation of the given data, called latent representation. Specifically, VAE [53, 54] is a generative model that can represent the continuous latent space of a given dataset. Using VAE, Noguchi and Inoue [43] presented a reconstruction methodology for generating steel microstructures from the given cooling rates. Their results demonstrated that the generated microstructures are in good agreement regarding morphology and quantitative aspects (e.g., ferrite volume fraction and grain size). Furthermore, Kim et al. [44] employed VAE to design the microstructure of dual-phase steel to satisfy target mechanical properties. They showed that their design framework enables controlling microstructural features in a continuous space within the latent space of VAE. Xu et al. [45] also exhibited that the morphology and the stochasticity of various microstructures (e.g., random fiber, particle, and ellipse) can be controlled using their trained VAE-based models. Meanwhile, a significant drawback of VAE is that the generated samples are often distorted and blurred, which can result in the quality of the samples not meeting desired standards [55, 56]. To address this problem, GANs have gained significant attention as an alternative for generating high-quality images with shaper and cleaner morphological features [57-59]. Fokina et al. [49] employed the StyleGAN [60] architecture to reconstruct microstructures while preserving the spatial distributions of original samples. Furthermore, Kench et al. [46] proposed a novel GAN architecture (i.e., SliceGAN) for synthesizing three-dimensional microstructures (e.g., polycrystalline grains, ceramic, fiber rods, etc.) using a single representative two-dimensional image. Gayon-Lombardo et al. [50] demonstrated a deep convolutional GAN for the reconstruction of multi-phase electrode microstructures preserving the spatial correlation from the original samples.

However, despite the extensive utilization of GANs in various research fields, GANs suffer from some critical problems which hinder their applications. One of the main challenges in using GANs is that the training process is unstable due to the nature of the loss function, which trains the generator against the discriminator in an adversarial manner [57, 61]. Due to this instability, GAN models often generate samples from a narrow range of probabilities (i.e., mode collapse). This phenomenon further increases the burden of adjusting the hyperparameters of models and the training process [62-64].

In recent days, diffusion-based generative models (DGMs) are becoming increasingly popular as a way to generate synthetic data. After the proposal of the diffusion model by Sohl-Dickstein [65], the model was further improved with the advanced probabilistic parameterization for the noising/denoising process [66, 67], which aided the training process and efficient application. Recent studies on DGMs have also shown that they can produce high-quality samples comparable to or even superior to those generated by GANs [66, 68-70]. In particular, a cascade of conditional diffusion models with a text encoder (i.e., Imagen) proposed by Google Research [70] achieved a state-of-the-art Fréchet inception distance (FID) score (7.27) in a zero-shot generation on the COCO dataset. The diffusion models also have demonstrated excellent generative performance in various research areas such as computer vision (e.g., semantic segmentation [71, 72] and super-resolution [73, 74]), natural language processing [75-77], test-to-image generation [69, 70], text-to-audio generation [78-80], and medical image reconstruction [81]. The recent studies by Lee and Yun [51] and Düreth et al. [82] also have demonstrated the applicability of DGM in sampling microstructures with various morphologies.

This paper proposes a novel DGM-based design framework for microstructure of multifunctional composites to pave the way for high-throughput material design. Owing to DGM and the advancement of DL architectures, it can be said that the era in which even inexperienced individuals can design materials solely based on data has become a tangible reality. To the best knowledge of the authors, this is the first study to develop the DGMs for ICME application, encompassing the three necessary steps in ICME-based material design: 1) characterization and reconstruction of microstructures, 2) analysis of material behavior to obtain material properties, and 3) inverse design of microstructures to achieve the desired performance (Figure 1). Like most approaches in materials engineering, we commence by performing experimental analyses to collect real-world data (i.e., micro-CT images) within an affordable range of trials (Figure 1a). In particular, the mechanoluminescence (ML) multifunctional composite with $SrAl_2O_4:Eu^{2+},Dy^{3+}$ (SAOED) and epoxy, which emits visible light intensity that is proportional to the magnitude of applied stresses [83], is chosen as the material of interest to validate the proposed design methodology. After data pre-processing, the underlying distribution of experimentally obtained microstructure data is learned by the unconditional DGM (Figure 1b) to generate morphologically equivalent microstructure samples. Since one of the crucial challenges in computational material design is creating a trustworthy large-scale database of microstructures, this step is essential before deriving the structure-property linkages in practice. Furthermore, it is also imperative to undertake a rigorous validation and comparison process to ensure the quality of the generated data regarding fidelity and diversity. Thus, the generated data's quality and morphological characteristics are evaluated with reliable evaluation metrics, including the FID score and the conventional statistical microstructure descriptors. A computational cost-efficient surrogate model (Figure 1c) is adopted in the next step to evaluate the generated microstructure samples in physical fields (e.g., stress fields). Although FEM has been a conventional numerical approach for solving various mechanical problems, its computational expense can become prohibitive when dealing with large data. Thus, a CNN-based surrogate model is built to quickly and accurately predict nonlinear material behavior with given microstructures and boundary conditions. Lastly, a conditional DGM (Figure 1d) with guidance is trained to generate microstructures with the desired material properties. Unlike conventional MCR methods [10-15] (e.g., statistical/morphological descriptor-based methods), a remarkable aspect of the proposed approach is that it does not require deep insights into microstructure morphologies. Instead, it involves implicitly learning the underlying data distribution, constructing a database through sampling, and enabling data-driven inverse design using generative/surrogate models.

The contents of this paper are organized as follows. Section 2 provides the formulations of generative models (i.e., unconditional/conditional DGMs) and the CNN-based surrogate modes for building the proposed data-driven microstructure design framework. Next, since the ML composite is considered a material of interest, section 3 briefly introduces this material and prepares training data (i.e., microstructure images and corresponding material properties) with the image processing and the multiscale analysis. After training the generative/surrogate models, many samples are created based on target material properties (i.e., volume fraction, elastic constant, and light sensitivity) to validate the design performance of the proposed framework. Section 4 discusses the quality of the generated microstructure samples in detail regarding the morphological similarities and the physical material properties. Finally, conclusions are summarized in Section 5, along with future research directions.

**Figure 1**. Overall flowchart for proposed design framework for composite microstructure using deep learned generative and surrogate models: (a) experimental analysis of the material (i.e., micro-CT imaging) and processing for FEM analysis of material properties, (b) unconditional DGM for microstructure reconstruction, (c) CNN-based surrogate model for prediction of material properties, and (d) conditional DGM for inverse design of microstructure

## 2. Models for the data-driven design of microstructure

To tailor the microstructure of multifunctional composites that suit particular purposes with desired properties, a three-step DL-assisted methodology is proposed in this study (Figure 1b-d). This section explains each step's necessary problem formulations and implementation details and the corresponding models.

2.1 Generative models for microstructure synthesis and design

As explained in the introduction, the proposed framework consists of two generative models, which are unconditional and conditional DGMs. Unlike the unconditional model to learn the original data distribution $p(\mathbf{x})$, the conditional model is required to learn the conditional distribution $p(\mathbf{x}|\mathbf{y})$, where $\mathbf{x}$ denotes the data and $\mathbf{y}$ is a set of given conditions, such as the data label. Since this study aims to build an inverse design framework for obtaining microstructures that have target material properties, the conditions (i.e., desired material properties) must be embedded and transferred to the model. In this section, the formulation of the diffusion-based generative models is first introduced. The method of guidance for generating data following conditional distribution is also described (section 2.1.1) as well as the implementation of multi-conditional embeddings (section 2.1.2).

2.1.1 Formulation of DGMs with guidance

The formulations of DGMs can currently be categorized into three types: denoising diffusion probabilistic models (DDPM) [84], score-based generative models (SGM) [85], and stochastic differential equations (SDE) [86]. Although there are slight variations in their formulations, all of these models rely on gradually adding random noise to data, followed by a denoising process to generate new samples. A brief overview of the DDPM-based formulation used in this study is provided here. For more detailed information, please refer to the references [66, 84, 87, 88].

DGMs are a type of latent variable model where the latent variables (i.e., noised samples) $\mathbf{z} = \{\mathbf{z}_t | t \in [0,1]\}$ follow a forward process that starts with the original data $\mathbf{x} \sim p(\mathbf{x})$. The forward process $q$ is the Markovian noising process (or Gaussian process) which progressively destroys the original data structure (Figure 2) as defined as

$$q(\mathbf{z}_t | \mathbf{x}) := N\big(\mathbf{z}_t; \sqrt{\bar{\alpha}_t}\mathbf{x}, (1 - \bar{\alpha}_t)\mathbf{I}\big) \tag{1}$$

where the noise schedule $\bar{\alpha}_t$ can be controlled by noise schedule parameter $\beta_t$ at each time step $t$ as follows.

$$\alpha_t := 1 - \beta_t \tag{2}$$

$$\bar{\alpha}_t := \prod_{i=0}^{t} \alpha_i \tag{3}$$

Additionally, a linear or cosine schedule [66, 68, 69, 89, 90] is typically used to define $\beta_t$ at each time step $t$, and the cosine schedule is employed for this study since it has been known to show better performance in image generation tasks [66]. DGMs are trained to reverse the forward process $q(\mathbf{z}_t \mid \mathbf{x})$ (i.e., denoise the latent variables $\mathbf{z_t}$), obtaining an estimate of the original data $\hat{\mathbf{x}}$. If a set of optional conditions $\mathbf{y}$ is added for the conditional reversing process, an estimate of $\hat{\mathbf{x}}(\mathbf{z}_t, \mathbf{y}, \beta_t)$ needs to be found. The optimization of the following squared error loss leads to the accomplishment of this task.

$$\mathbb{E}_{\mathbf{x},\mathbf{y},\epsilon,t}\left[\left\|\hat{\mathbf{x}}\left(\sqrt{\bar{\alpha}_t} + \epsilon\sqrt{(1-\bar{\alpha}_t)}, \mathbf{y}\right) - \mathbf{x}\right\|_2^2\right] \quad (4)$$

where $\epsilon$ is the noise component for parameterizing the model [66, 84]. This approach of transforming generation into denoising can be supported by the optimization of a weighted variational lower bound (VLB) on the data log-likelihood within the probabilistic modeling framework [66, 68, 84, 87] or by considering it as a type of denoising score matching [85, 86].

Furthermore, to generate a specific sample that meets certain criteria during the denoising process, it is necessary to regulate the diversity of the generated output through conditioning. For instance, generative models like GANs and flow-based models possess a unique capability of performing truncated or low-temperature sampling by restricting the range of noise inputs to the model during sampling. While this technique can reduce the diversity of generated samples, it can improve the quality of each sample. However, applying truncated or low-temperature sampling to diffusion models is not effective, as scaling or reducing the variance of Gaussian noise during the reverse process can result in blurry samples [68, 91].

To address this problem, the classifier-free guidance [91] approach is adopted in this study for conditioning the denoising process. Unlike the classifier guidance, this approach does not require any pre-trained model for adding guidance terms for generating samples [68].

Instead, a single DGM is trained to optimize conditional and unconditional objectives by randomly omitting (i.e., dropping) a set of conditions **y** with the following parameterization.

$$\tilde{\epsilon}(\mathbf{z}_t, \mathbf{y}) = w\hat{\epsilon}(\mathbf{z}_t, \mathbf{y}) + (1-w)\hat{\epsilon}(\mathbf{z}_t) \tag{5}$$

The first and second terms on the right-hand side are the conditional and unconditional predictions of $\epsilon$, respectively. The parameter $w$ is for controlling the guidance, where $w = 0$ leads to an unconditional prediction of $\epsilon$. In addition, following previous references [70, 91], the omitting probability of 10% for guidance has been chosen to jointly optimize the unconditional and conditional objectives for DGMs in this study.

**Figure 2**. Schematic of forward noising process in DGMs for microstructure image synthesis

2.1.2 Implementation of DGMs

Although DGMs are highly flexible and can be used with any DL model, U-Net architecture [92] is mainly adopted for the iterative denoising process of DGMs since its input dimension is the same as the output dimension. To enhance the performance of DGMs with U-Net architecture, a significant number of parametric studies [66, 68, 70, 85, 90, 93] have been conducted in recent years. Among various architectural improvements for unconditional/conditional DGMs, it has been found that incorporating residual blocks and attention modules (i.e., self-attention and cross-attention) are beneficial for high-quality data synthesis. In this regard, the architecture of U-Net with residual block and multi-head attention is employed in this study, as can be seen in Figure 3. Figure 4 illustrates the residual block used in the U-Net, which consists of two sub-blocks (surrounded by the blue box in the figure). Each sub-block consists of three components in sequence: batch normalization, ReLU [94] activation, and convolution. The skip connection is used for learning concerning the input of the residual block. Furthermore, the multi-head attention, which is used in transformer [95, 96], is applied at some resolutions, as shown in Figure 3, with the U-Net layer self-attention and cross-attention (for conditional embeddings (Section 2.1.3)). In addition, the attention modules are known to improve the performance of DL models as they enable the model to concentrate on the most pertinent parts of the input, which can be helpful in various data modalities such as material engineering [97, 98] and state-of-the-art text-to-image generation [70, 99]. The down/up-sampling blocks (red/green arrows in Figure 3) are composed of two residual blocks with the $2 \times 2$ max pooling (for down-sampling) and $2 \times 2$ transposed convolution (for up-sampling). Other parameters for DGMs, including the number of discrete diffusion steps, are shown in Table 1. PyTorch library [100] is used to implement all the models in this study, including the DGMs and the U-Net. The models are trained using Nvidia RTX A6000 Graphics Processing Units (GPUs) and the Adam optimizer with a learning rate of $10^{-4}$.

**Figure 3**. Architecture of U-Net for DGMs with residual block and multi-head attention

**Figure 4**. Residual block composed of two sub-blocks in DGMs

**Table 1**. Parameters used in unconditional/conditional DGMs for designing microstructure of multifunctional composites in this study

| Overall DGM parameters | |
|---|---|
| Image size | 128 |
| Condition drop probability | 0.1 |
| Number of forward (i.e., noising) steps | 1000 |
| Conditional embedding dimension | 512 |
| Noising schedule | Cosine |
| Optimizer | Adam optimizer; learning rate of $10^{-4}$; 10000 warm-up steps |
| Batch size | 32 |
| U-Net (in DGMs) parameters | |
| Dimension | 128 |
| Dimension multipliers | 1,2,4,8 |
| Layer attentions | False, False, True, True |
| Layer cross-attentions | False, False, True, True |
| Number of Resnet blocks | 2 |
| Number of attention heads | 8 |
| Dimension of attention head | 64 |

2.1.3 Multi-conditional embeddings for microstructure synthesis

As introduced in the previous section, passing the conditional signal to the DGMs is necessary for conditional data synthesis. In this study, the following three conditional design parameters are considered for example volume fraction of ML particles $V_f$, effective elastic modulus $\tilde{C}_{11}$, and ML composites' light intensity change ratio (LICR) (detailed explanations are provided in the next section). For the generation of microstructure samples that meet these conditions, the design parameters need to be transformed into an appropriate space that the models can assess. The value of each design parameter is projected into a high-dimensional embedding space using the positional encoding approach [101] (Appendix) to accomplish this. This approach utilizes sine and cosine functions with different periods to enable smooth interpolation between the values of each design parameter. In addition, this method has been effectively used to incorporate material information (e.g., a fractional amount of constituent elements) with DL models in the previous literature [102]. Then, the corresponding conditional embeddings are pooled and added into the time step embedding in DGMs [68]. Furthermore, the model is conditioned on the design parameters via cross-attention, concatenating the embeddings with the key-value pairs of self-attention layers in U-Net (Figure 5), a conditioning method used by Imagen [70].

**Figure 5**. Conditioning DGMs on the microstructure design parameters

It is worth noting that although the model can be conditioned on the embeddings using cross-attention with the U-Net self-attention layers, as mentioned earlier, the self-attention map between the design parameters can also be considered. It means that the self-attention values of the design parameters can be passed to the concatenation block in Figure 5, allowing for the incorporation of the relative importance of each parameter in advance. To achieve this, a suitable transformer encoder should be trained to encode the design parameters into an appropriate space where semantic information can be easily assessed for a specific task, such as analyzing material behaviors [98, 102]. However, this task is not conducted as it is beyond the scope of this study. The results obtained for the example case (i.e., design of ML composites (Section 4)) are considered acceptable to demonstrate the effectiveness of the proposed framework. In addition, the authors suggest that this could be a potential research direction in the future.

2.2 Surrogate models for prediction of stress and strain fields

To establish reliable microstructure-property linkages, a substantial amount of analysis (experimental or numerical) needs to be conducted, considering different microstructure morphologies and boundary conditions. However, even numerical techniques such as FEM are often not computationally feasible for analyzing material behavior with many microstructure samples, especially when considering nonlinear material behavior (e.g., plasticity). Thus, a cost-efficient surrogate model with a U-Net based architecture is employed in this study to fully take advantage of the large-scale dataset of microstructure available with the DGMs.

Although the U-Net architecture was originally developed for medical image segmentation [92], it has been demonstrated to capture the underlying features of various types of data effectively. Recently, Bhaduri et al. [52] also showed that the U-Net architecture is effective in predicting the stress field in the 2-D microstructure of fiber-reinforced composites with unidirectional loading conditions. The U-Net model used in this study is a modified version of the original U-Net architecture, which also differs from the model used by Bhaduri et al. [52], as shown in Figure 6. Similar to the U-Net architecture in the proposed DGMs (Section 2.1.1), two residual blocks (where each consists of two sub-blocks, as shown in Figure 4) are employed to deepen the network, whereas the original structure consists of two single convolution blocks. Furthermore, multi-head self-attention with the number of attention heads 8 is applied at a resolution of $8 \times 8$. It is also worth noting that the input is a one-channel binary phase microstructure image, and the output is a one-channel image of the stress field, unlike the standard U-Net with three input/output channels.

As explained in the next section, the purpose of this surrogate model is to predict stress components in a specific direction ($\sigma_{11}$) with relatively small strain imposed for obtaining effective elastic modulus and to predict the von Mises stress ($\sigma_v$) field with a relatively large strain imposed to incorporate plastic behavior. Thus, two separate U-Nets (with the same architecture for simplicity) are trained to acquire the $\sigma_{11}$ field and the $\sigma_v$ field, respectively. The weights of the proposed architecture are trained by minimizing the loss function, which is defined as the mean squared error between the predicted stress map and the ground-truth stress map (i.e., FEM simulation results) in the training dataset.

**Figure 6**. Architecture of U-Net with residual block and multi-head attention for surrogate models to predict stress fields

# 3. Experimental and numerical Analysis of multifunctional composite materials for building a training dataset

## 3.1. SAOED particulate composite materials

As previously stated, the material chosen to validate the proposed DGM-based material design framework is an ML composite composed of SAOED and epoxy. SAOED is a persistent luminescent material (PLM) composed of strontium aluminate codoped with lanthanide ions, such as europium and dysprosium ions [103]. In addition, PLMs have emerged as promising multifunctional materials with applications extending beyond traditional luminous materials to various uses. Notably, SAOED exhibits intense light emission under mechanical deformations, as illustrated in Figure 7. This light emission is visible in broad daylight with the naked eye. Consequently, SAOED has been effectively employed in developing structural health monitoring systems capable of detecting cracks, stress concentrations, impacts, and film pressure [104-107].

PLMs have recently attracted attention from various research fields due to their potential application in various mechano-optical devices and non-destructive evaluation. Numerous studies have demonstrated the feasibility of utilizing SAOED to visualize and measure stress distribution. Significant advancements have been made in the technical implementation of SAOED as a full-field non-destructive tool for measuring stress and strain, particularly in epoxy/SAOED composites such as sensing films and adhesive laminae [108, 109]. However, several challenges must be addressed to harness these materials' potential fully. One such challenge is enhancing the sensitivity of the materials, which is critical for their widespread application, as it allows for more accurate and precise detection [110-112]. SAOED particles are commonly incorporated into epoxy resin, facilitating the transfer of external stresses to the particles. Moreover, it has been found that deviatoric stresses, which cause distortion or deformation, are more effective in triggering the release of trapped charges in ML phenomena compared to hydrostatic stresses, which implies that it is reasonable to utilize the von Mises stress as a function of the amount of light intensity [113].

**Figure 7**. Schematic image of mechanoluminescence (ML) material

## 3.2. CT image characterization and reconstruction

Before micro-CT imaging, ML composites are prepared by mixing SAOED powder (LumiNova®, G-300M) with epoxy resin (Smooth-On Inc., EpoxAcast™ 690), with the SOED power comprising 70% wt of the total composite. Considering the specific gravities of constituents, the volume fraction of powder takes up 41.62%. To ensure uniform transmission of X-rays, the ML composite is molded into a cylindrical shape with a 1mm diameter. A 6C-beam line at the Pohang Accelerator Laboratory (PAL, South Korea) is employed to obtain the CT images. The camera's field of view (FOV) measures $16.6 \times 14 \text{mm}^2$, and the pixel size is 6.5μm. Considering the powder size, a 10x microscope magnification is applied, resulting in a pixel size of 0.65μm for the images. The whole process for micro-CT imaging is depicted in Figure 8.

**Figure 8**. Schematic process of micro-CT scanning and scanned volume and obtained RVE element

Using the acquired micro-CT images, multi-step image processing is performed with Simepleware® software, summarized in Figure 9. First, median and recursive Gaussian filters are sequentially applied to the original image for denoising purposes. Subsequently, the watershed algorithm is employed to extract distinct particles. The final reconstructed image appears binary concerning its constituents. Readers can refer to our previous work for more details regarding micro-CT imaging and processing [26].

**Figure 9**. Micro-CT image processing: (a) raw image, (b) watershed algorithm, and (c) binary image

### 3.3. Microstructure modeling and analysis with FEM

The binary pixels should be converted into solid elements for FE simulation with the given microstructure image samples. Figure 10a shows nodes and connectivity are created according to Abaqus/Standard input format. The resulting FEM model is generated, as depicted in Figure 10b. The image size is cropped to 128×128 before converting to the FEM model, resulting in 16384 elements (the original image size is 300×300). To analyze the behavior at the micro-scale, FE model constructed using each microstructure sample is treated as RVE. Then, appropriate boundary conditions need to be defined. The RVE should possess features allowing neighboring RVEs to fit together in deformed and un-deformed states. Consequently, the boundary conditions for the RVE should be periodic to preserve the continuity of displacements, strains, and stresses across each RVE [114]. For implementation in FEM, the boundary conditions can be expressed as linear constraints, and implemented as multipoint constraints.

$$u_i^- - u_i^+ - \Delta L_x \varepsilon_{i1} - \Delta L_y \varepsilon_{i2} = 0 \tag{6}$$

Here, $u_i^-$ represents the displacement of the node on the slave region (−) and $u_i^+$ denotes the displacement of the node on the master region (+), which corresponds to the other region of (−). $\Delta L_i$ is the relative distance between two nodes. By employing the following equation, the nodes on the 2D RVE can be grouped according to their location, such as edge and vertex nodes. This grouping is based on the differing relative distances depending on the group and helps prevent nodes from being over-constrained.

Then, the RVE models are subjected to strain loading, expressed by $\Delta \boldsymbol{\varepsilon} = \Delta \varepsilon \boldsymbol{\psi}$ with $\boldsymbol{\psi} = \boldsymbol{e_1} \otimes \boldsymbol{e_1}$ under periodic boundary conditions. The constituents assume isotropic material. The properties of epoxy resin EpoxAcast$^{TM}$ are $E = 3.94$ GPa and $\nu = 0.3$, while SAOED particles are $E = 102$ GPa and $\nu = 0.23$. Additionally, plastic behavior is incorporated at the matrix region (i.e., epoxy resin) to represent the nonlinear behavior of the ML composite material. SAOED particles are assumed to exhibit only linear behavior with relatively high mechanical properties. The plastic behavior is realized using the $J_2$ flow rule while considering isotropic hardening [115]. The material parameters related to plasticity are set such that the initial yield stress is 20 MPa and the hardening slope is 2.1 GPa, corresponding to the supplier's specification. The loading strain is applied as 0.2% in the tensile direction. Through the FE analysis, the von Mises stress contour of the microstructure is depicted in Figure 10c. Because of the difference in material properties between constituents, the particle region shows a higher stress level than the matrix region. Furthermore, the stress concentration occurs at the adjacent regions between particles and the matrix.

**Figure 10**. FEM modeling from a reconstructed image: (a) binary image, (b) FEM model, and (c) von Mises stress contour

**Figure 11**. Randomly selected microstructure images and corresponding von Mises stress contours

Figure 11 shows the obtained stress fields of randomly selected samples of ML microstructure. Due to the different morphologies of the samples, the maximum and minimum stress levels vary. Based on the stress-strain distribution of composite materials, we can evaluate the mechanical properties of the composite materials. Moreover, the light intensity is defined by the volume-average stress of the particle region, which can be converted into ML sensitivity. Those indicators will be utilized as target design parameters in the design problem in the following section. The effective mechanical behavior of each region is obtained through a volume-averaging scheme written as follows.

$$\tilde{\sigma}_i = \frac{\int_V \sigma_i dV}{\int_V dV}$$

$$\tilde{\varepsilon}_i = \frac{\int_V \varepsilon_i dV}{\int_V dV} \qquad (7)$$

Using the volume-averaging scheme, the effective elastic modulus ($\tilde{C}_{11}$) of composites can also be obtained based on the volume-averaged stress as

$$\tilde{C}_{11} = \frac{\tilde{\sigma}_{11}}{\tilde{\varepsilon}_{11}} \qquad (8)$$

However, it should be noted that the FEM analysis is conducted on a model subjected to constraints (zero strain) in other directions (i.e., except for 11-direction). It is because $\tilde{\varepsilon}_{22}$ and $\tilde{\varepsilon}_{33}$ can influence the result of $\tilde{C}_{11}$ [116]. On the other hand, the reason why the unconstraint conditions are applied for von Mises stress calculation is to take into account Poisson's effect.

Next, to estimate the light intensity, we adopt a theoretical model that considers the light emission mechanism, including a de-trapping of charge carriers and their subsequent recombination [117]. They validated this model with experimental testing by showing the light intensity change ratio (LICR) as a function of the particle size and tensile loading speed. Because the stress-strain curves show nonlinear regardless of strain rate, LICR–strain curves represent nonlinear behavior. The LICR can be expressed in exponential form as follows:

$$\text{LICR} = c \frac{\exp(at) - at - 1}{t} \tag{9}$$

Here, $c$ and $a$ are the material parameters derived from Boltzmann's statistical formula and vary depending on the particle size and strain rate. $t$ is the experimental time, which is assumed to be five seconds. In this simulation, we define $c = 400$ and $a = \tilde{\sigma}_v^{particle}/400$ [MPa] with consideration of reference microstructure images. $\tilde{\sigma}_v^{particle}$ is the volume-averaged von Mises stress in particle region. Then, the $\tilde{C}_{11}$ and LICR are employed as conditional design parameters along with the $V_f$ as introduced in Section 2.

## 4. Results and Discussion

4.1 Generated microstructure samples

To train the unconditional DGM for the generation of microstructure samples, the microstructure images obtained from micro-CT imaging (section 2.2) are used as a training dataset, along with a 4-fold augmentation that includes 90°, 180°, and 270° rotations (resulting in a total of 1200 microstructure images, i.e., 300 images multiplied by 4). The model is then trained for 60.000 iterations with the given training dataset. After training, 5,760 samples are generated using the conditional DGM. The randomly selected generated ML microstructure samples using the unconditional DGM with the experimentally obtained micro-CT data are shown in Figure 12. Here, it can be observed that the unconditional DGM can produce microstructure samples that closely resemble the original samples in terms of visual similarity. To quantify the performance of DGM in generating visually similar samples, the FID score (5.70) is calculated as presented in Table 2. According to previous research [51], the FID scores for DGM-generated binary microstructure samples were reported to be around 2-20. Additionally, Google Research [70] achieved an FID score of 7.27 on the COCO datasets using the DGM approach. Therefore, it can be said that the FID score obtained in this study is comparable to those achieved by state-of-the-art generative models. As depicted in Figure 13, the two-point correlation function $S_2(r)$ [13] and the lineal path function $L(r)$ [118] are also evaluated to examine the morphological discrepancies between the original and generated ML microstructure samples (where the shaded envelope represents the deviations of the functions). The values of the discrepancy [119] in $S_2(r)$ and $L(r)$ between the original and DGM-generated samples are 3.56% and 1.46% (Table 2), respectively. Furthermore, the results indicate that the distribution of particle volume fractions is well preserved in the generated samples, which is supported by the similarity in the correlation functions (in terms of mean and deviation) of the original and generated samples when the distance between randomly selected points (r) is near zero. Overall, the evaluation results for the generated ML samples indicate a good agreement between the original and DGM-generated samples.

Furthermore, to demonstrate that the evaluated samples are not selectively chosen (i.e., cherry-picked) for reducing the discrepancy between the generated and original samples, the original samples that have the minimum mean squared pixel error $\varepsilon_{pixel}$ with respect to the randomly selected generated samples are shown in Figure 14. The $\varepsilon_{pixel}$ for each pair of original and generated samples with $N(\text{row}) \times M(\text{column})$ size can be computed with the pixel values of original and generated samples ($y_{ij}$ and $\hat{y}_{ij}$, respectively) as follows.

$$\varepsilon_{pixel} = \frac{1}{NM} \sum_{i=1}^{N} \sum_{j=1}^{M} (y_{ij} - \hat{y}_{ij})^2 \tag{10}$$

As shown in the figure, it is evident that the generated and original samples exhibit different phase distributions with black pixels ($y_{ij} = 0$) for matrix and white pixels ($y_{ij} = 1$) for ML particles. This observation indicates that the generated samples are not mere replicas of the training dataset. In other words, the trained DGM can produce novel morphologically equivalent microstructure samples that can be utilized for subsequent analysis, such as material properties analysis. To provide a more detailed analysis of the discrepancies between the original and generated samples, the distribution of minimum $\varepsilon_{pixel}$ is depicted in Figure 15. Here, there are cases where $\varepsilon_{pixel}$ is near zero, indicating data regurgitation (i.e., data copying [120]), a common problem in generative models. However, the model itself can create genuinely new data samples (but have equivalent morphological characteristics with the original samples) as shown in the distribution with $\varepsilon_{pixel} > 0$. Additionally, it is noteworthy that certain samples, such as the third and fourth samples in Figure 14, exhibit partially similar morphology to the original training dataset. Nevertheless, these samples possess distinct overall morphologies. Previous literature research has also documented the capability of DGMs to generate images that bear a partial resemblance but exhibit notable differences in their general characteristics while retaining contextual consistency [51]. The results also demonstrate that the developed DGM does not suffer from mode collapse, a frequently observed problem in GANs [62]. The remarkable aspect is that the provided results are obtained without the need for any fine-tuning or truncation tricks to obtain novel but equivalent data samples, highlighting the easy accessibility and extensive applicability of the DGM across diverse domains and disciplines in the future.

**Figure 12**. Microstructure samples of ML composites: (a) experimental data obtained using micro-CT and image processing, (b) generated data using the trained unconditional DGM

**Figure 13**. Comparison of two-point correlation functions and lineal path functions between the original microstructure samples and the generated microstructure samples

**Table 2**. FID scores and a discrepancy between the spatial correlation functions for DGM-generated and original samples

| FID score | $\varepsilon_{S_2}$ (%) | $\varepsilon_L$ (%) |
|---|---|---|
| 5.70 | 3.56% | 1.46% |

**Figure 14**. Randomly selected samples generated with the unconditional DGM and the original samples that are closest in pixel space

**Figure 15**. Distribution of minimum $\varepsilon_{pixel}$ (in pixel space) of the generated samples with unconditional DGM with respect to the closest original samples in pixel space

4.2 Predicted stress fields with U-Net

To train U-Net surrogate models for the prediction of stress fields ($\sigma_{11}$ and $\sigma_v$), only one-third of the generated samples (1,920 samples out of 5,760 samples) are used, and the corresponding stress fields are obtained using the FEM analysis (Section 2.3). The constructed data pairs (i.e., microstructure-stress fields) are then used as a training dataset, with 10% of the dataset used for validation. In addition, the trained U-Net is used to predict the stress fields for the remaining samples, allowing the construction of microstructure-properties data pairs for training the conditional DGM (section 4.3) with a small computational cost. The model is then trained for 1,000 epochs using a batch size of 32, with the mean squared error between the true and predicted stress fields as the training objective. The training loss and validation loss for the U-Net surrogate models are shown in Figure 16. The results demonstrate that the models perform well within the training and validation datasets as the training progresses.

**Figure 16**. Loss function values during training and validation of U-Net surrogate models for (a) prediction of $\sigma_{11}$ and (b) prediction of $\sigma_v$

The comparison between the true stress fields obtained through FE analysis and the predicted stress fields using the U-Net surrogate model for a microstructure sample (Figure 17a) is depicted in Figure 17. With a small strain, 0.02%, applied in 11 direction, $\sigma_{11}$ fields are predicted to obtain the volume-averaged stress $\tilde{\sigma}_{11}$ as well as the effective elastic modulus $\tilde{C}_{11}$. Compared to the true $\sigma_{11}$ field (Figure 17b), the predicted $\sigma_{11}$ field (Figure 17c) shows similar locations of stress concentration with a relatively small error (Figure 17d). Meanwhile, for obtaining LICR (as described in Section 3.3) for each microstructure sample, the $\sigma_v$ field needs to be obtained. By applying a relatively large strain, 0.2%, imposed in 11 direction, $\sigma_v$ fields are predicted to obtain the volume-averaged von Mises stress in the particle regions ($\tilde{\sigma}_v^{particle}$). In comparison to the true $\sigma_v$ field (Figure 17e), the predicted $\sigma_v$ field (Figure 17f) also exhibits similar locations of stress concentration while with a relatively small error (Figure 17g). To assess the accuracy of the predicted stress fields for each sample, the mean absolute error, maximum absolute error, and normalized root mean squared error (RMSE) are computed as shown in Table 3 with the following definitions:

$$\text{Mean absolute error} = \frac{1}{NM} \sum_{i=1}^{N} \sum_{j=1}^{M} \left| y_{ij}^{\sigma_k} - \hat{y}_{ij}^{\sigma_k} \right| \tag{11}$$

$$\text{Maximum absolute error} = \text{Max}_{ij} \left[ \left[ \left| y_{ij}^{\sigma_k} - \hat{y}_{ij}^{\sigma_k} \right| \right]_i \right]_j, \quad i = 1, \ldots, N, \tag{12}$$

$$j = 1, \ldots, M$$

$$\text{Normalized RMSE} = \frac{1}{F_{Norm}} \times \sqrt{\frac{\sum_{i=1}^{N} \sum_{j}^{M} \left( y_{ij}^{\sigma_k} - \hat{y}_{ij}^{\sigma_k} \right)^2}{NM}} \tag{13}$$

where $y_{ij}^{\sigma_k}$ and $\hat{y}_{ij}^{\sigma_k}$ denote the reference stress and predicted stress, respectively. The normalizing factor $F_{Norm}$ represents the difference between the maximum and minimum reference stress values. In this study, the range of true $\sigma_{11}$ is $[-9.58, 32.62]$ and the range of true $\sigma_v$ is $[0.19\ 298.57]$, resulting in $F_{Norm}$ values of 42.2. and 293.38, respectively. It is worth noting that the evaluation metrics are computed for each sample, and Table 3 shows the mean and standard deviation of the values after evaluating all the microstructure samples. For instance, the value of $3.37 \times 10^{-2}$ and the value of $6.38 \times 10^{-1}$ for the mean absolute error for $\sigma_{11}$ and $\sigma_v$ represent that the mean absolute error of whole evaluated samples. Furthermore, the mean values of the maximum absolute error are 1.32 and 20.68. Considering the $F_{Norm}$ values for each type of stress, the relative percentage amounts of these maximum absolute error values are 3.13% and 7.04%. Overall, the evaluation metrics indicate a high degree of concordance between the predicted values of stress fields with U-Net surrogate models and the corresponding true stress fields. In this regard, the surrogate models can be used to evaluate $\tilde{\sigma}_{11}$ and $\tilde{\sigma}_v^{particle}$ for obtaining properties of ML composites (e.g., $\tilde{C}_{11}$ and LICR).

**Figure 17**. Predicted stress fields for using the trained surrogate models (U-Nets): (a) input microstructure sample, (b) true $\sigma_{11}$, (c) **predicted $\sigma_{11}$**, (d) error map for $\sigma_{11}$, (e) true $\sigma_v$, (f) predicted $\sigma_v$, (g) error map for $\sigma_v$

**Table 3**. Accuracy metrics for the predicted stress fields with surrogate models (U-Nets)

| Mean absolute error ($\sigma_{11}$) | Maximum absolute error ($\sigma_{11}$) | Normalized RMSE ($\sigma_{11}$) | Mean absolute error ($\sigma_v$) | Maximum absolute error ($\sigma_v$) | Normalized RMSE ($\sigma_v$) |
|---|---|---|---|---|---|
| $3.37 \times 10^{-2} \pm 2.13 \times 10^{-2}$ | $1.32 \pm 1.14$ | $5.28 \times 10^{-3} \pm 3.96 \times 10^{-3}$ | $6.38 \times 10^{-1} \pm 3.76 \times 10^{-1}$ | $20.68 \pm 18.14$ | $5.34 \times 10^{-3} \pm 8.04 \times 10^{-3}$ |

4.3 Inverse-designed microstructures samples

After constructing microstructure-property data pairs using the results obtained from U-Net surrogate models (5,760 data pairs), 4,800 data pairs are selected as the training dataset for the conditional DGM. The remaining property data from these pairs are then used as input target design parameters to validate the conditional DGM for the inverse design of ML composites' microstructure. It is necessary to ensure that the input design parameters are physically realistic and accurately represent the range of ML composite properties present in the dataset. Significantly, the content of target material properties are $[0.3256, 0.5056]$, $[8.0135, 13.4983]$, $[3.4069, 13.0019]$ for $V_f$, $\tilde{C}_{11}$, and LICR, respectively. The conditional DGM (section 2.1.2) is then trained for 100.000 iterations using the constructed training dataset. Figure 18 shows the comparison between the target material properties and the properties of the corresponding generated samples. It can be observed that the properties of the generated samples exhibit a similar trend to that of the target material properties. The error distribution for each property in Figure 18 also demonstrates that the errors are centered around zero, indicating that the conditional DGM can generate samples that align with the input target properties (i.e., design criteria). To quantitatively evaluate the performance, the mean absolute error, maximum absolute error, and coefficient of determination (i.e., $R^2$ score) between the target properties and the properties of generated samples are computed as shown in Table 4. The high values of $R^2$ scores ($\geq 0.95$) indicate a strong correlation between the target properties and generated samples.

Several samples with varying target material properties are illustrated in Figure 19 to demonstrate the developed unconditional DGM's capability for designing ML composites' microstructures. It is evident that as $V_f$ increase (from (a) to (d)), the ML particles coarsen, and the fraction of the matrix region decreases. Along with the rise in $V_f$, the effective elastic modulus ($\tilde{C}_{11}$) increases since the modulus of ML particles is much higher than that of the matrix. On the other hand, the value of LICR does not always increase as $V_f$ increases as shown in Figure 19b and c. It is because LICR is calculated using the volume-averaged von Mises stress within the particle region. Indeed, an increase in $V_f$ generally leads to higher average brightness in the ML composites. However, it is crucial to consider the sensitivity of light emission, as it plays a significant role in determining the suitability of these composites for various applications. Furthermore, the distribution and shape of ML particles are also significant descriptors contributing to the variation of LICR. The nonlinearity of the material behavior resulting from plasticity in the matrix is also influenced by redistribution of stress within particle regions. In other words, the problem of designing ML composites in this study is a nonlinear multi-objective design problem. The remarkable aspect is that the developed conditional DGM model has successfully addressed this problem using a data-driven approach assisted by the unconditional DGM and surrogate models. Moreover, since the proposed framework is based on a data-driven approach, the methodology can be easily extended to other design problems, including the design of different materials. Using a limited amount of experimental data (e.g., micro-CT images), the proposed DGM-based framework enables the inverse design of material microstructures without encountering critical issues such as model collapse and non-convergence [57, 62].

On the other hand, the proposed methodology requires further validation with highly nonlinear problems. A more sophisticated model becomes necessary when designing materials that involve many design parameters and if learning their correlations is challenging. In particular, one possible solution is to pre-train a suitable transformer encoder for embedding the parameters in a well-interpolated embedding space, which can facilitate the learning of complex relationships between input values. Subsequently, the conditional DGM can utilize the self-attention values to generate new samples by cross-attention, concatenating the embeddings. Furthermore, it is crucial to validate whether the generated microstructures can be fabricated in the real world, considering their manufacturability. Nevertheless, the proposed methodology can serve as a foundational framework for high-throughput materials design by offering practical microstructure reconstruction and inverse design capabilities.

**Figure 18**. Target material properties vs material properties of generated samples using conditional DGM with the error distribution around the mean: (a) $V_f$, (b) $\widetilde{C}_{11}$, and (c) LICR

**Figure 19**. Variation of generated ML microstructure samples obtained using the conditional DGM with respect to the variation of design parameters

**Table 4**. Accuracy metrics for the generated samples with condition DGMs

| Property | Mean absolute error | Maximum absolute error | $R^2$ score |
|---|---|---|---|
| $V_f$ | $1.91 \times 10^{-3}$ | $2.18 \times 10^{-2}$ | 0.98 |
| $\tilde{C}_{11}$ [GPa] | $2.19 \times 10^{-1}$ | $9.97 \times 10^{-1}$ | 0.96 |
| LICR | $3.44 \times 10^{-1}$ | 1.98 | 0.95 |

## 5. Conclusions

This study proposes a novel data-driven framework for designing microstructures of multifunctional composites using unconditional/conditional DGMs and CNN-based surrogate models (i.e., U-Nets). By utilizing the unconditional DGM to generate equivalent microstructure samples and the surrogate model to predict stress fields, the training dataset can be established for microstructure-property linkage with faster computational speed, facilitating the data-driven inverse design of microstructure. As a means of validating the proposed methodology, the design of ML composites microstructures was examined, taking into account three target material properties ($V_f$, $\tilde{C}_{11}$, and LICR) for an illustrative case study. The experimentally obtained micro-CT images of ML composites were employed as the initial reference to account for real-world situations. Utilizing the unconditional DGM, more microstructure samples were generated by learning the underlying distribution of original micro-CT images to enrich the training dataset for inverse design. The low FID score and the discrepancies observed between the original and generated samples in terms of the two-point correlation function and the lineal path function demonstrated the potential of the unconditional DGM for microstructure reconstruction. The material properties of the generated samples were then obtained using the predicted stress fields with U-Net surrogate models. The accuracy metrics, including the mean absolute error and normalized RMSE, indicated that the surrogate models could predict stress fields with a small error range. Finally, the conditional DGM for the inverse design of ML composite microstructures was trained using the data pairs obtained using unconditional DGM and surrogate models. The results show that the inversely designed microstructures with conditional DGM align closely with the target material properties. It is also worth noting that the design of ML composites is a nonlinear multi-objective problem, and the proposed DGM-based framework successfully addressed it.

Since the proposed framework encompasses the common ICME process, which involves the reconstruction of microstructures and the analysis of material properties using a data-driven approach, it can be directly applied to designing other materials for specific purposes. For instance, various materials, including composites and polycrystalline microstructures of alloys, can be considered for real-world ICME applications, utilizing the proposed DGM-based framework for high-throughput material design. For future recommendations, enhancing the model architecture for inverse material design, such as integrating a pre-trained transformer encoder, could be explored to fine-tune the models for designing materials with highly nonlinear behavior. Additionally, it is crucial to consider manufacturability by incorporating suitable methods, such as providing manufacturing parameters to the DGMs to capture the distribution of feasible manufacturing conditions.

## Appendix

To utilize the design parameters ($V_f$, $\tilde{C}_{11}$ and LICR) as conditional signals to guide the DGMs to generate samples that meet specific target values, the values of the design parameters are transformed into a high-dimensional space using the positional encoding approach [95, 101]. Similar to the compositionally restricted attention-based network (CrabNet) presented by Want et al. [102, 121], a value of the design parameter can be transformed into the output vector $E_p$ of the embedding dimension $d$ with the following formula.

$$E_p^{(2i)} = \sin\left(\frac{p}{n^{(2i/d)}}\right) \tag{14}$$

$$E_p^{(2i+1)} = \cos\left(\frac{p}{n^{(2i/d)}}\right) \tag{15}$$

where $i$ denotes a column index ($0 \leq i < d/2$) and $p$ is the value of interest. The values of the design parameters are normalized on a linear scale with a resolution of 0.0001 within their minimum and maximum range prior to the encoding process. Then, the values are encoded into a smooth interpolated high-dimensional space and normalized within $[-1, 1]$ (due to the sine and cosine functions), as shown in Figure 20.

**Figure 20.** Visualization of encoding within normalized value of design parameter

**Acknowledgments**

This material is based upon work supported by the Air Force Office of Scientific Research under award number FA2386-22-1-4001 and the Institute of Engineering Research at Seoul National University. The authors are grateful for their support. This work was also supported by the BK21 Program funded by the Ministry of Education (MOE, Korea) and National Research Foundation of Korea (NRF-4199990513944).